\documentclass[10pt,a4paper]{article}
\usepackage[utf8]{inputenc}
\usepackage[margin=2.5cm]{geometry}
\usepackage{graphicx}
\usepackage{amsmath}
\usepackage{amssymb}
\usepackage{cite}
\usepackage{hyperref}
\usepackage{float}
\usepackage{lineno}
\usepackage{caption}
\usepackage{subcaption}
\usepackage{booktabs}
\usepackage{multirow}
\usepackage{markdown}
\usepackage{pdfpages}
\usepackage{listings}
\usepackage{url}
\usepackage{tabularx}
\usepackage[T1]{fontenc}
\usepackage{xcolor}

\title{Hallucination Detector: A hybrid LLM and Semantic Scholar tool calling for detecting hallucination in scientific literature on AtomGPT.org}
\author{Harichandana Neralla$^{1}$, Jaehyung Lee$^{2}$, Aldo H. Romero$^{4}$, Kamal Choudhary$^{2,3,*}$\\[0.8em]
\small$^{1}$Hopkins Extreme Materials Institute, Johns Hopkins University, Baltimore, MD 21218, USA\\
\small$^{2}$Department of Materials Science and Engineering, Johns Hopkins University, Baltimore, MD 21218, USA\\
\small$^{3}$Department of Electrical and Computer Engineering, Johns Hopkins University, Baltimore, MD 21218, USA\\
\small$^{4}$Department of Physics and Astronomy, West Virginia University, Morgantown, WV 26506, USA\\[0.6em]
\small$^{*}$Corresponding author: \href{mailto:kchoudh2@jhu.edu}{kchoudh2@jhu.edu}}

\date{\today}

\begin{document}
\maketitle

\begin{abstract}

Large language models are now commonly used as partners in scientific writing, and this shift has brought a subtler type of failure: made‑up references. Fabricated authors, bogus DOIs, wrongly assigned identifiers, and citations that merge elements from multiple genuine articles are now being inserted into manuscripts at a volume that traditional peer review was never meant to handle. Recent audits reveal that such references have already slipped through the review process and made their way into the published literature, including leading journals and conferences. Automated verification that operates at the speed and scale of modern content production has therefore become a necessary safeguard rather than a convenience. This work presents and evaluates the AtomGPT reference checker (\url{https://atomgpt.org/hallucination_detector}), an open, web-accessible tool that verifies citations against the scholarly literature by combining large-language-model field extraction with structured retrieval from Semantic Scholar. For each reference, the tool extracts the bibliographic fields, retrieves the closest matching real papers, and scores the agreement across title, authorship, and venue to produce a graded judgment of
whether a citation is trustworthy, partially supported, or likely fabricated. We benchmark the tool against an externally curated set of confirmed hallucinated citations from accepted NeurIPS 2025 papers and find that it reliably flags the great majority of them. A per-field analysis explains the behavior: fabricated references are most commonly uncovered when the listed authors do not correspond to any actual publication, even if the title appears credible, while similarity in titles alone is a weak indicator, and similarity in venues is an even weaker one. We characterize the small number of missed
cases in which a fabricated citation closely resembles a single real paper and position the tool as a lightweight, drop-in component for editorial pipelines, submission systems, and review platforms, where catching fabricated references before publication is increasingly essential to preserving research integrity.
\end{abstract}

\section{Introduction}

Citations are fundamental to the integrity of the scientific record. Each reference signals that a verifiable source exists and supports the statement it accompanies. When a cited source does not exist or cannot be verified, readers, reviewers, and future researchers lose the ability to trace the origin of the information and to evaluate the evidence independently.

Large language models (LLMs) are now routinely used as partners in academic writing, and this shift has introduced a significant new failure mode: reference hallucination, in which an LLM generates citations that appear authentic, often including plausible authors, journal names, publication years, and other bibliographic details, but do not correspond to any real publication. Such fabricated references can be difficult to detect because they closely resemble legitimate citations. Recent studies have begun to quantify how prevalent this behavior is. Walters and Wilder found that a substantial fraction of the bibliographic citations generated by general-purpose chatbots were either fabricated or contained significant bibliographic errors, with the frequency varying considerably across models and model generations~\cite{walters2023}. In the medical domain, where the impact of fabricated references is particularly serious, Bhattacharyya et al. found that only a small fraction of the citations in chatbot-generated text were both genuine and correct~\cite{bhattacharyya2023}, and Athaluri et al. found that a large share of generated DOIs failed to resolve~\cite{athaluri2023}. Similar patterns recur across disciplines, including geography~\cite{day2023} and law, where hallucinated case citations appear in a majority of model responses~\cite{dahl2024} and persist even in commercial retrieval-augmented legal tools marketed as hallucination-free~\cite{magesh2025}. Chelli et al. compared chatbots on systematic-review references and observed high fabrication rates with notable variation between systems~\cite{chelli2024}.

These fabrications are no longer confined to informal or low-quality outlets; they now enter the published literature. A large-scale audit of 2.5 million biomedical papers and roughly 97 million references found that the share of papers containing at least one fabricated reference rose more than tenfold in three years, from about one in 2{,}828 papers in 2023 to one in 277 in early 2026, with the sharpest increase coinciding with the widespread availability of LLM writing tools~\cite{topaz2026}. A parallel finding in machine learning supplies the benchmark used in this paper: a study by GPTZero scanned 4{,}841 accepted NeurIPS 2025 papers and uncovered 100 verified hallucinated citations spread across more than fifty of them, references that had each cleared review by three to five expert referees before appearing in the official proceedings~\cite{gptzero2026}. These fabrications ranged from entirely invented papers with non-existent authors to subtler corruptions, such as real identifiers pointing to unrelated work or titles and fields recombined from several genuine papers. Ansari subsequently carried out a manual review of this same set of 100 citations, assigning each to one of five failure-mode categories and finding that the vast majority were entirely fabricated, with a sizable secondary portion showing partial corruption of an otherwise genuine reference~\cite{ansari2026}. That such citations survive peer review at a top-tier venue demonstrates that human scrutiny alone cannot keep pace with the volume and verisimilitude of machine-generated references.

Detecting these citations at scale requires automated tooling, and the natural design is retrieval-grounded: rather than asking a model what it believes, one retrieves the bibliographic details for each reference and verifies them against an authoritative index of actual publications. This paper presents and evaluates one such tool, the reference checker exposed by AtomGPT~\cite{choudhary2024,lee2026agapi}, an open, web-accessible platform for scientific AI that also hosts a suite of generative and analysis tools for materials design~\cite{choudhary2025diffractgpt,choudhary2025microscopygpt,choudhary2025jarvis,lee2026hackathon}. The tool combines two components: an open-weight LLM (gpt-oss-20b) that parses a free-form citation into structured fields (title, authors, venue, year), and a retrieval-and-match stage that queries Semantic Scholar~\cite{kinney2023}, a free academic search service indexing over two hundred million papers for the closest real papers and scores the agreement on each field. The individual field similarities are merged into one overall composite score, which is then compared to a threshold to assign one of three graded statuses: \textit{verified}, \textit{partial}, or \textit{likely hallucinated}. This architecture is deliberately lightweight: it does not need model fine-tuning, provides an interpretable breakdown by field explaining why a citation was flagged, and operates on a continuously updated scholarly index.

We evaluate this tool on the GPTZero NeurIPS 2025 benchmark, treating the 100 confirmed hallucinations as a recall test on known positives, and we analyze not only how many it catches but \emph{why}. Our contributions are threefold. First, we provide an independent benchmark of the AtomGPT reference checker on an externally curated, expert-verified set of fabricated citations, reporting its detection rate and the small number of failures. Second, we decompose the tool's composite score into its title, author, and venue components and show that detection is driven overwhelmingly by author mismatch, while plausible titles make title similarity a weak standalone signal; this provides a mechanistic account of how retrieval-grounded checkers succeed and where they fail. Third, we document the tool's practical limitations, including the run-to-run variability introduced by its LLM extraction stage and the structural inability of a positives-only benchmark to estimate precision, and we outline how a balanced evaluation and multi-source verification would address them. Together, these results position the AtomGPT reference checker both as a usable safeguard for editorial workflows and as a case study in what makes automated citation verification work.

% \section{Related Work}

Fabricated citations are one specific instance of the wider problem of hallucination in large language models. Huang et al.\ offer an extensive overview of this area, classifying hallucinations into a taxonomy according to their forms, root causes, detection techniques, and mitigation approaches~\cite{huang2023survey}.

A growing body of work aims to detect these failures. In the context of scientific writing, Vangala et al. introduced HalluMat, a multi-stage framework for detecting factual hallucinations in materials science that combines intrinsic verification, retrieval from multiple sources, and contradiction analysis~\cite{hallumat2025}; because HalluMat targets incorrect scientific claims, such as erroneous material properties, rather than fabricated citations, it addresses a distinct challenge and complements the citation verification considered here. Methodologically, Agrawal et al. suggested exploiting a model's own inconsistencies in generated author lists as a black-box signal for detecting hallucinated references~\cite{agrawal2024}, while Aljamaan et al. proposed a Reference Hallucination Score and showed that retrieval-augmented systems significantly surpass unguided chatbots~\cite{aljamaan2024}. Several systems verify references directly by extracting their fields and matching them against scholarly databases. The closest tool to the one studied here is CheckIfExist, which parses each citation, queries multiple scholarly indices, and computes a multi-field similarity to classify references, treating fabricated authorship as a primary diagnostic signal~\cite{abbonato2026}; its architecture is nearly identical to AtomGPT's reference checker, the principal difference being that CheckIfExist cross-validates across several databases whereas the tool evaluated here matches against a single index behind an LLM extraction stage. Notably, CheckIfExist was presented without a quantitative benchmark on an external hallucination set, a gap that the present evaluation helps fill. On the benchmarking side, Yuan et al. introduced CiteAudit, a benchmark for verifying scientific references in the LLM era that distinguishes genuine citations from fabricated and misattributed ones~\cite{yuan2026citeaudit}.

The verification approach used here builds on established ideas. Combining title, author, and venue similarities into a single match score is an instance of probabilistic record linkage, whose theoretical foundations were laid by Fellegi and Sunter~\cite{fellegi1969} and later extended with learnable per-field similarity measures by Bilenko and Mooney~\cite{bilenko2003}. Complementary scholarly indices such as arXiv~\cite{arxivapi} and Crossref~\cite{crossref} provide additional preprint and DOI coverage beyond the Semantic Scholar backend used by the tool evaluated here.

\section{Methods}

Our objective is twofold: to measure how reliably the AtomGPT reference checker identifies citations that are already known to be fabricated, and to understand the per-field mechanism behind its decisions. We therefore structure the evaluation as a citation-level recall test on known positives. Because every reference we submit has already been verified as fake, the question the study addresses is narrow and well-defined: when the tool is given a citation that is known to be fabricated, how consistently does it flag it? This design is intentional. It isolates the tool's detection behavior from the harder, separate task of locating hallucinations within a complete reference list, and it allows every result to be checked against an external answer key rather than against our own judgment. The consequence, which we make explicit throughout, is that the study measures recall and the mechanism behind it, not precision; we return to this boundary in Section~\ref{sec:limitations}.

The benchmark is the publicly released GPTZero NeurIPS 2025 dataset~\cite{gptzero2026}: 100 confirmed hallucinated citations drawn from 53 distinct accepted papers out of the 4{,}841 papers GPTZero scanned. Each citation was first flagged by GPTZero's automated checker and then manually verified by a member of their team, which is what makes the set usable as ground truth: the fabrication label on each citation is externally assigned and human-checked, not inferred by us. The fabrications include placeholder author names (e.g., ``John Doe and Jane Smith''), fake DOIs, fabricated arXiv identifiers, real arXiv IDs pointing to unrelated papers, ``Frankenstein'' references that combine fields from multiple real papers, and invented venues. The unit of analysis is the individual citation, not the paper.

\begin{figure}[H]
    \centering
    \includegraphics[width=0.85\linewidth]{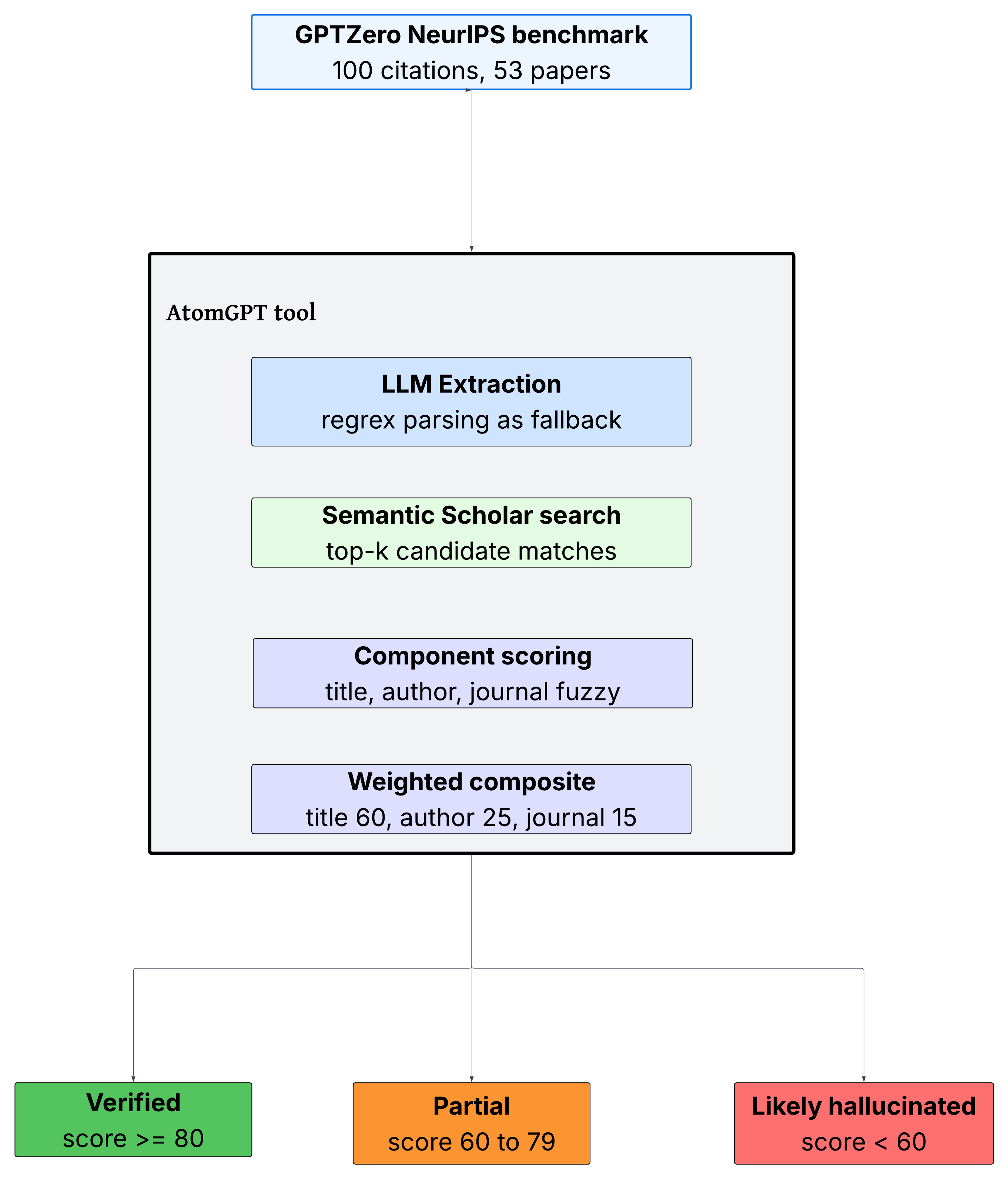}
    \caption{Evaluation pipeline. The GPTZero NeurIPS 2025 benchmark supplies
    100 confirmed hallucinated citations (across 53 distinct papers) as input to
    the AtomGPT reference checker. Inside the tool, LLM-based extraction (with a
    regular-expression parser as fallback) parses each citation into title,
    authors, journal, and year; Semantic Scholar is queried for the top-$k$
    candidate papers; per-field title, author, and journal similarities are
    computed against the best match and combined into a weighted composite
    (title $60\%$, authors $25\%$, journal $15\%$). A threshold on the composite
    assigns each citation to \textit{verified} ($\geq 80$), \textit{partial}
    ($60$--$79$), or \textit{likely hallucinated} ($<60$).}
    \label{fig:pipeline}
\end{figure}

Figure~\ref{fig:pipeline} summarizes the end-to-end evaluation. Each of the 100 citations is submitted individually to the AtomGPT reference checker through its web interface. Internally, the tool extracts the reference into structured fields, queries Semantic Scholar for the closest real papers, computes per-field similarity scores against the best match, combines them into a single composite score, and emits a status label.

\begin{figure}[H]
    \centering
    \includegraphics[width=0.85\linewidth]{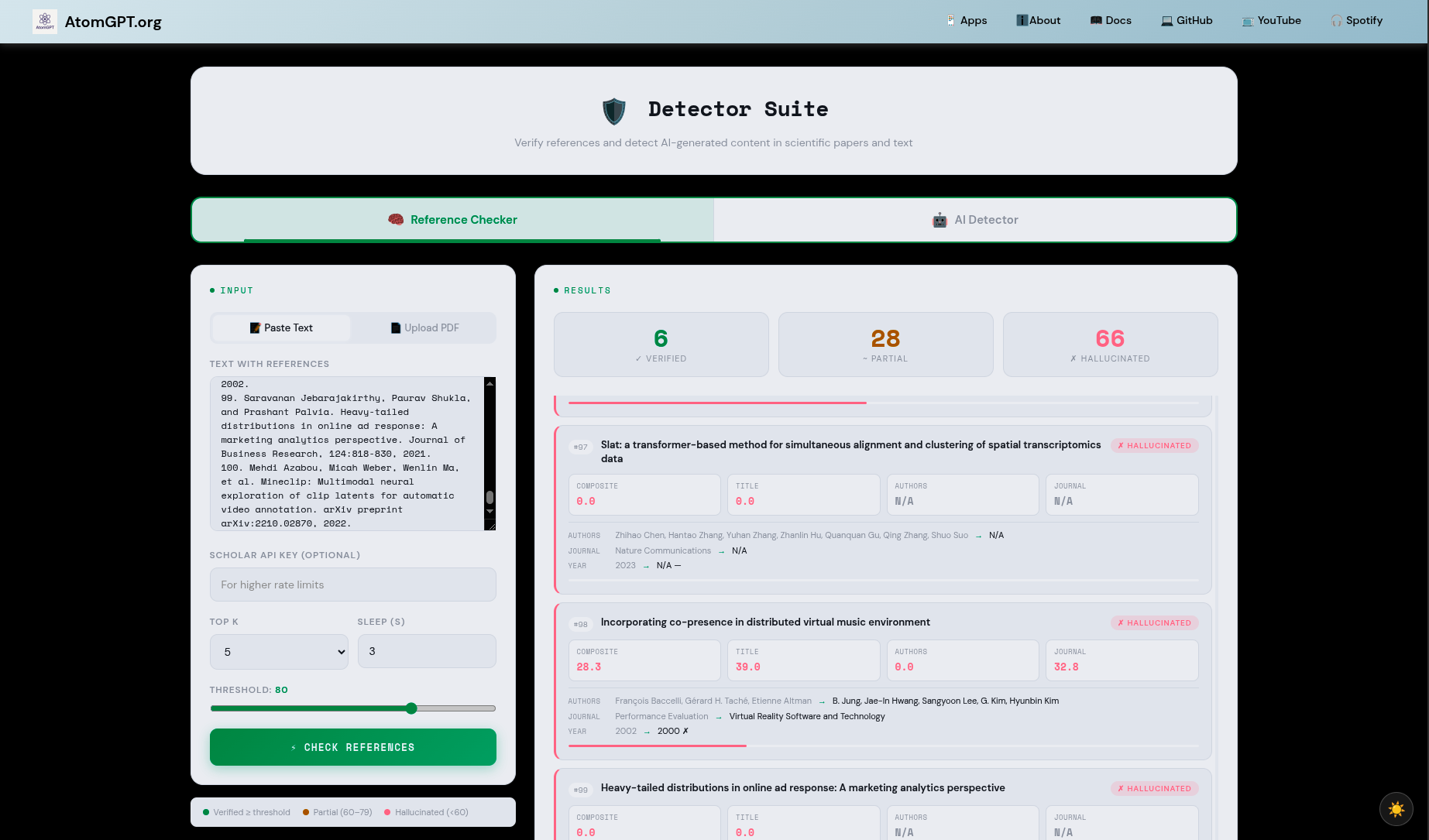}
    \caption{Snapshot of the AtomGPT reference-checking web interface showing a
    representative result: the submitted citation, the best-matching real paper
    retrieved from Semantic Scholar, the per-field title, author, and journal
    similarity scores, and the resulting composite score and status label.}
    \label{fig:tool-snapshot}
\end{figure}

We enter each citation into the AtomGPT reference checker via its web interface individually and then export the results for each reference. For each
citation, the tool reports a composite score, the per-field title, author, and
journal similarity scores, a Boolean year match, the best-matching candidate
paper, and a status label. Figure~\ref{fig:tool-snapshot} shows the interface with a representative result. We keep the individual scores for each field, rather than just the final label, because they enable us to justify \emph{why} a citation is assigned a particular label; the analysis in Section~\ref{sec:results} depends on this more detailed breakdown.

The system first uses gpt-oss-20b, an open-weight LLM released by OpenAI~\cite{openai2025gptoss}, to parse each citation into structured bibliographic fields, including the title, authors, journal, and publication year. If this extraction step fails or no reference is identified, a local regular-expression parser is used as a fallback. Regardless of the extraction method, the resulting bibliographic fields are processed by the same verification pipeline. Because the LLM-based extraction is stochastic, identical citation strings may produce slightly different field assignments across independent runs. These differences can affect the generated search query, the publication retrieved from Semantic Scholar, and, in some cases, the final classification. By contrast, all subsequent stages of the pipeline, including similarity scoring and threshold-based classification, are fully deterministic. The implications of this stochastic component for reproducibility are discussed in Section~\ref{sec:limitations}. The extraction model is a configurable component of the tool and may change in future versions; the results reported here reflect the model in use at the time of evaluation.

Each candidate returned by Semantic Scholar is
scored on three fuzzy field similarities, each on a $0$--$100$ scale: title similarity ($s_t$) between the citation title and the candidate title, author
similarity ($s_a$) as the fraction of cited authors whose surnames appear in the candidate's author list, and journal similarity ($s_j$) between the cited venue
and the candidate venue. Each similarity thus captures a distinct way in which a citation can either match or diverge from the closest real paper, and this is what later enables us to assign the detection to a particular field. The composite is a weighted sum with title weighted $60\%$, authors $25\%$, and journal $15\%$:
\begin{equation}
S = 0.60\,s_t + 0.25\,s_a + 0.15\,s_j .
\end{equation}
When a field is absent from the extraction (no authors or no venue parsed), the
corresponding term substitutes the title score rather than zero, so a missing field does not penalize the composite the way a mismatched field does. This
design choice matters for interpretation: a low author or journal score reflects
a genuine mismatch against a real paper, whereas a missing field is simply
excluded, which is why we report each component over the subset of citations for
which it was actually extracted. The year is recorded as a Boolean match but does not enter the composite. The reported score for each citation is the maximum $S$ over the top-$k$ candidates, i.e., the tool gives the citation the benefit of its single best match.

The status is a threshold on the composite:
\textit{verified} for $S \geq 80$, \textit{partial} for $60 \leq S < 80$, and
\textit{likely hallucinated} for $S < 60$. These three bands are the tool's graded verdict, ranging from a citation it judges trustworthy to one it judges
fabricated, and they are the labels we compare against the benchmark.

The scholarly source is Semantic Scholar~\cite{kinney2023},
queried through its top-$k$ paper-search API. Because Semantic Scholar is a live,
rate-limited service, throttled or empty responses occasionally yield no candidates and therefore a zero score; this is a second, run-dependent source of
variation discussed in Section~\ref{sec:limitations}.

\section{Results}
\label{sec:results}

\begin{figure}[H]
    \centering
    \includegraphics[width=0.65\linewidth]{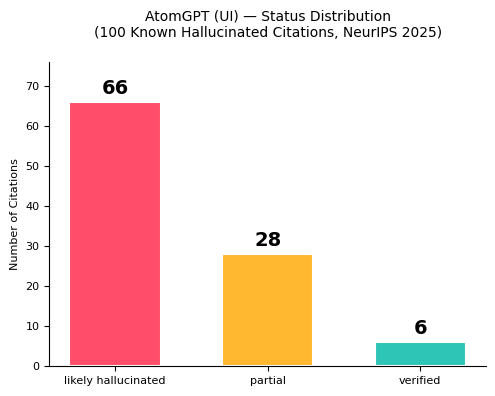}
    \caption{AtomGPT detection status across the 100 known hallucinated
    citations from the GPTZero NeurIPS 2025 benchmark: 66 \textit{likely
    hallucinated}, 28 \textit{partial}, and 6 \textit{verified}. Treating
    \textit{likely hallucinated} and \textit{partial} as detected gives a $94\%$
    detection rate.}
    \label{fig:status}
\end{figure}

The first and most basic question is how many of the known fabrications the tool
catches. Of the 100 confirmed hallucinated citations, AtomGPT classified 66 as
\textit{likely hallucinated}, 28 as \textit{partial}, and 6 as \textit{verified}
(Figure~\ref{fig:status}). We treat both \textit{likely hallucinated} and
\textit{partial} as detections since both represent the tool declining to
confirm the citation as real, while reserving \textit{verified} for cases where
the tool actively endorsed a fabricated reference. Under this reading, the tool
flagged 94 of 100 citations, a $94\%$ detection rate, and the 6 \textit{verified}
citations are the true misses: fabrications the tool would have passed as
genuine. Under a stricter reading that counts only \textit{likely hallucinated}
as a confident positive, the rate is $66\%$; the 28 \textit{partial} citations are better understood as the tool withholding endorsement than as confident
detections, and the two readings ($66\%$ and $94\%$) bracket the tool's behavior at its strict and lenient operating points. In both cases, the tool effectively performs its task; the rest of the results section details the behavior that drives this performance.

The detection rate counts decisions; the composite score shows how confident those decisions were and whether the threshold is well placed. We therefore
examine the distribution of the composite across all 100 citations. The mean
composite was 46.7 (median 51.6, standard deviation 26.5), with the 25th and 75th
percentiles at 36.4 and 64.4 and a full range of $0$ to $100$. We report the mean
and median together because the gap between them indicates skew: both sit below
the \textit{partial} threshold of 60, which is the expected location for a set
that is almost entirely fabricated, and the spread up to 100 reflects a minority
of citations whose surface form happens to match a real paper closely. The most
informative cut is the mean composite within each status band, which rises
monotonically (33.3 for \textit{likely hallucinated}, 69.4 for \textit{partial},
88.5 for \textit{verified}). This monotonic separation is what tells us the score
is not assigning labels arbitrarily: citations the tool is most confident are
fake score lowest, and the score increases smoothly toward the cases it endorses,
so the three bands correspond to genuinely different degrees of match rather than
to an arbitrary partition of a single undifferentiated pile.

\begin{figure}[H]
\centering
\includegraphics[width=0.95\linewidth]{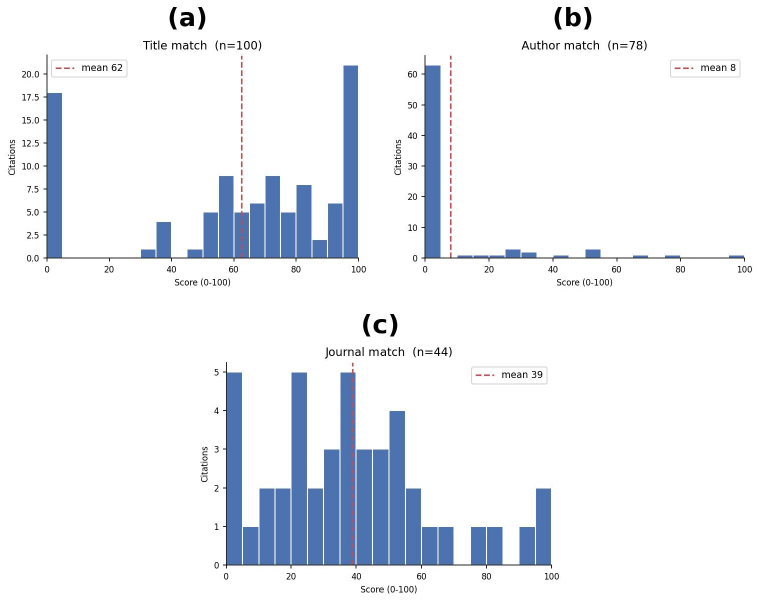}
\caption{\textbf{Per-field similarity score distributions across the scored
citations.} Each panel shows one component of the composite score, reported over
the subset of citations for which that field was extracted, so a low score always
reflects a genuine mismatch rather than a missing field.
\textbf{(a) Title similarity} over the 100 scored citations (mean 62.4): despite
every citation being fabricated, title scores cluster high with a heavy concentration near 100, showing that fabricated titles closely resemble those of
real papers.
\textbf{(b) Author similarity} over the 78 citations with extracted authors (mean
8.0): scores collapse toward zero, as the cited authors match no real paper in the
overwhelming majority of cases; this is the dominant signal driving detection.
\textbf{(c) Journal similarity} over the 44 citations with extracted venues (mean
38.9): scores are spread across the range with no clear separation, consistent
with journal similarity carrying the least discriminative signal of the three
components. The contrast between the high title scores in (a) and the
near-zero author scores in (b) is the mechanism by which the tool exposes
fabricated references.}
\label{fig:component-panels}
\end{figure}

A detection rate tells us \emph{that} the tool works; it does not tell us
\emph{why}, or \emph{when} it fails. To recover the mechanism, we decompose the
composite into its three component scores and ask which field actually drives the
verdict. Because author and journal scores exist only when those fields were
extracted, the components have differing sample sizes (title $n=100$, author
$n=78$, journal $n=44$); we report each over its available subset rather than
imputing zeros, so that a low score always means a real mismatch and never a
missing field. Figure~\ref{fig:component-panels} shows the distribution of each
component.

Author similarity is the primary discriminator. The mean author score climbs
sharply across the status bands, from
3.5 for \textit{likely hallucinated} to 8.1 for \textit{partial} and 60.4 for
\textit{verified}, and among the \textit{likely hallucinated} citations that had authors extracted,
45 of 47 scored below 30 (Figure~\ref{fig:component-panels}b). The interpretation is direct: the author score is near
zero for almost every citation the tool flags, and it is the one field that
separates the flagged citations from the few it endorses. In practical terms,
fabricated references pair a plausible-sounding title with an author list that
matches no real paper, and it is the author check that exposes them. This is
consistent with GPTZero's own qualitative notes describing fabricated and placeholder author names.

Title similarity is high even for fakes, which is why a title-only check would
be insufficient. It is substantial across
every band (mean 45.9 for \textit{likely hallucinated}, 93.6 for \textit{partial}, 98.8 for
\textit{verified}; overall mean 62.4 across all 100 citations), and Figure~\ref{fig:component-panels}a shows a heavy
concentration near 100. Since every citation in this set is fabricated, the high
title scores tell us that the fabricated titles closely resemble those of real
papers; a detector relying on title alone would therefore endorse most of them.
Detection works only because the author check overrides a convincing title.

Journal similarity carries little signal. The journal-score means do not
separate the bands in a
consistent direction (34.3 for \textit{likely hallucinated}, 49.3 for
\textit{partial}, 22.0 for \textit{verified}), and the field was extractable for
fewer than half the citations (Figure~\ref{fig:component-panels}c). We read this as evidence that the $15\%$ journal
weight adds little discriminative information on this dataset, both because venue
strings are often missing or generic and because, where present, they do not
track the fabrication.

For completeness we record whether the cited year matched the
year of the best-matching real paper. The year agreed in only 12 of 100
citations and disagreed in 70, with 18 citations carrying no extractable year
(Table~\ref{tab:year-match}). The predominance of mismatches is consistent with
fabrication, but we treat the year as descriptive context only: it does not enter
the composite and therefore plays no role in the tool's decision.

\begin{table}[H]
    \centering
    \begin{tabular}{lrr}
        \toprule
        \textbf{Year match} & \textbf{Count} & \textbf{Share (\%)} \\
        \midrule
        False (mismatch)      & 70 & 70.0 \\
        True (match)          & 12 & 12.0 \\
        Not available         & 18 & 18.0 \\
        \midrule
        \textbf{Total}        & 100 & 100.0 \\
        \bottomrule
    \end{tabular}
    \caption{Year-match outcomes across the 100 citations. The cited year
    disagreed with the matched paper in 70 cases and matched in only 12 (18 had
    no extractable year). The year is recorded for description only and does not
    enter the composite score.}
    \label{tab:year-match}
\end{table}

Finally we examine the six citations the tool missed, classified as
\textit{verified}, because they reveal its failure modes. Table~\ref{tab:false-neg}
lists all six. They share one feature without exception: a near-perfect title
score (92.9 to 100), confirming that every missed fabrication carried a title
closely matching a real paper. Beyond the title, however, the six split into two
distinct mechanisms. In two cases the cited authors genuinely matched the
best-matching real paper (author scores of 75 and 100), so a real title and real
authors together pushed the composite above the \textit{verified} threshold; these
are the ``Frankenstein'' fabrications that recombine fields from genuine papers.
In the remaining four, the author field was either weakly matched (a score of
33.3) or not extracted at all. The two with no extracted authors are the more
instructive: under the composite's missing-field rule, any author or venue term that cannot be extracted is substituted with the title score instead of being downgraded. As a result, a fabricated citation whose author list cannot be parsed is effectively evaluated only on how plausible its title appears and ends up being accepted. The tool's errors thus stem from the same title-and-author weighting scheme that underlies its successes: it fails either when a fabricated entry happens, by chance, to align with a real paper on both of these heavily weighted fields, or when a missing field lets a compelling title pass without being challenged.

\begin{table}[H]
    \centering
    \begin{tabular}{cccc}
        \toprule
        \textbf{Title} & \textbf{Author} & \textbf{Journal} & \textbf{Composite} \\
        \midrule
        100.0 & 33.3  & ---  & 83.3  \\
        100.0 & 75.0  & 14.0 & 80.9  \\
        100.0 & 100.0 & ---  & 100.0 \\
        100.0 & ---   & ---  & 100.0 \\
         92.9 & ---   & 30.0 & 83.4  \\
        100.0 & 33.3  & ---  & 83.3  \\
        \midrule
        \textbf{98.8} & \textbf{60.4} & \textbf{22.0} & \textbf{88.5} \\
        \bottomrule
    \end{tabular}
    \caption{Per-field scores for the six fabricated citations the tool
    misclassified as \textit{verified}. Dashes denote fields the extractor did not
    parse, which the composite replaces with the title score rather than
    penalizing. All six have near-perfect title scores; only two have genuinely
    high author matches, while two have no author score at all. The bottom row
    gives the means (author and journal means are over the non-missing entries:
    $n=4$ and $n=2$ respectively).}
    \label{tab:false-neg}
\end{table}

\section{Discussion}
\label{discussion}

Our findings can be read against two recent studies of the same underlying
problem. The first is a manual analysis of the identical 100-citation set by
Ansari~\cite{ansari2026}, who hand-coded each fabrication into a five-category
failure-mode taxonomy: total fabrication (66\%), partial attribute corruption
(27\%), identifier hijacking (4\%), placeholder hallucination (2\%), and semantic
hallucination (1\%). The distribution is striking next to ours. Ansari's two
dominant categories, wholesale fabrication and partial corruption of an otherwise
real reference, account for 93\% of the set in proportions (66\% and 27\%) that
closely track our tool's automated split of 66 \textit{likely hallucinated} and
28 \textit{partial}. The two analyses are methodologically independent: Ansari
classifies the \emph{nature} of each fabrication by expert human reading, whereas
our tool assigns a status from automated field matching against Semantic Scholar,
with no knowledge of any taxonomy. That an unsupervised matching procedure
recovers nearly the same coarse structure as careful manual coding is broadly
consistent with that structure being real: the bulk of these hallucinations are
clean inventions that fail every field check, while a substantial minority retain
enough genuine attributes to score partially. We do not claim a one-to-one
correspondence at the level of individual citations: Ansari reports that every
hallucination in fact exhibits \emph{compound} failure modes, layering several
deception strategies at once, which is precisely the condition under which our
tool's six false negatives arise.

The second point of comparison is architectural. CheckIfExist~\cite{abbonato2026}
is, to our knowledge, the closest existing tool to the AtomGPT reference checker:
it likewise extracts a citation's fields, queries scholarly indices, computes a
multi-field similarity, and treats fabricated authorship as a primary diagnostic
signal. Two systems arrived independently at the same conclusion our per-field
analysis reaches, that authorship is the decisive field, which strengthens the
claim that author mismatch is the core mechanism of retrieval-grounded citation
verification rather than an artifact of one implementation. The principal design
difference is breadth of evidence: CheckIfExist cross-validates across several
databases (Crossref, Semantic Scholar, OpenAlex), whereas the tool evaluated here
matches against a single index. As CheckIfExist was presented without a
quantitative benchmark on an external hallucination set, the present study also
supplies a missing piece of evidence for this class of tool: a detection rate
measured against expert-verified, real-world fabrications.

\subsection{Limitations}
\label{sec:limitations}
Several boundaries constrain what this evaluation can claim.

The benchmark contains only positives. Every citation in the GPTZero
set is a confirmed fabrication, so the study measures recall, the fraction of
known fakes the tool catches, but cannot estimate precision or the
false-positive rate. A tool that flagged \emph{every} citation would score
perfectly on this benchmark. Characterizing how often the tool wrongly flags
genuine references requires a balanced set of real and fabricated citations, which
we did not have; this is the single most important gap and the natural subject of
a follow-up study.

The pipeline is non-deterministic. The LLM extraction stage is
stochastic, so identical inputs can produce slightly different field parses, and
hence different matches and occasionally different final statuses, across runs.
The downstream scoring and thresholding are deterministic, but the variation
introduced upstream means a single run is not exactly reproducible. Reported
counts should therefore be read as representative of a typical run rather than as
fixed constants, and a fuller characterization would report the mean and spread
of the status distribution over repeated runs.

Verification relies on a single, live source. The tool matches against
Semantic Scholar alone. A genuine citation absent from that index, or temporarily
unreturned because the rate-limited service throttled the request, yields no
candidate and a zero score, which would register as a (false) hallucination in a
mixed setting. Single-source dependence also means the tool inherits any coverage
gaps of that index.

The journal field is frequently unavailable. A venue was extracted for
only 44 of the 100 citations, and where present its similarity did not separate
the status classes consistently. The journal component therefore contributes
little reliable signal in practice, and the effective decision rests almost
entirely on title and author agreement.

The scoring weights are misaligned with the discriminative signal.
The composite weights the title at $60\%$ and authorship at $25\%$, yet our
per-field analysis shows the reverse ordering of usefulness: authorship is the
decisive discriminator, while title similarity is high even for fabrications and
is therefore a weak standalone signal. Compounding this, the missing-field rule
substitutes the title score for an unextracted author or venue term, so a
fabrication with an unparseable author list is effectively graded on its title
alone. Both choices favor the least reliable field, and three of the six false
negatives arise directly from a strong title overriding a weak or missing author
signal. Re-weighting toward authorship, treating a low or missing author score as
a partial-status trigger rather than substituting the title, or applying an
explicit author veto are natural mitigations; because each is a change only to the
deterministic scoring stage, the present data could be used to test them directly,
which we leave to future work.

The benchmark consists of general machine-learning
papers rather than materials-science papers. Because the reference checker matches
against a domain-general scholarly index, we expect its behavior to transfer; however,
we do not evaluate domain-specific performance here.

\section{Conclusion}

We evaluated the AtomGPT reference checker on an externally curated set of 100
confirmed hallucinated citations that had survived peer review at NeurIPS 2025,
and found that it flags 94 of them, missing only six. A per-field analysis
explained the result mechanistically: detection is driven overwhelmingly by
author mismatch, since fabricated references tend to carry plausible titles but
authors that correspond to no real paper, and the few misses are exactly those
fabrications that coincidentally resemble a single real paper on both title and
authorship. The tool's automated status distribution closely mirrors an
independent manual taxonomy of the same citations, and its reliance on author
agreement matches the design of the nearest comparable system, two convergences
that suggest the underlying approach is sound rather than incidental.

Three directions follow naturally. The most important is a balanced evaluation on
a mixture of genuine and fabricated citations, which would yield the precision and
false-positive rate this study cannot. The second is multi-source verification:
cross-checking candidates across several scholarly indices, as some contemporary
tools do, would directly attack the false-negative mode we identified, since a
fabrication that matches one index is less likely to match several. The third is a
characterization of run-to-run variability, reporting the distribution of outcomes
over repeated runs so that the tool's behavior can be stated with quantified
confidence. Together these would extend the present recall study into a complete,
deployment-ready account of automated citation verification.

\section*{Data and Code Availability}
The AtomGPT reference checker evaluated here is openly accessible through its web
interface at \url{https://atomgpt.org}. The benchmark of 100 confirmed
hallucinated citations is from the publicly released GPTZero NeurIPS 2025
investigation~\cite{gptzero2026}. The evaluation notebook, per-citation outputs
(composite and per-field scores, matched candidates, and status labels), and
figure-generation code underlying the figures and tables in this paper are
publicly available at \url{https://github.com/harineralla/AtomGPT-Hallucination-Detector}.

\section*{Conflict of Interest}
On behalf of all authors, the corresponding author states that there is no conflict of interest.

\section*{Acknowledgements}
West Virginia University effort was supported by the West Virginia Higher Education Policy Commission through the Research Challenge Grant Program 2022 (Award No. RCG 23-007) and by the National Science Foundation under Award No. OAC-2513657.

\bibliography{references}

\end{document}